\documentclass[sort&compress]{elsarticle}
\usepackage{amsmath,mathtools,amssymb}
\usepackage{graphicx}   
\usepackage{epstopdf}
\usepackage{subfigure}
\usepackage{color}
\usepackage{verbatim} 
\usepackage{booktabs}
\usepackage{bm}   
\usepackage{dutchcal}
\usepackage[left=3cm,right=3cm,top=2.5cm,bottom=2.5cm]{geometry}  
\usepackage{lineno,hyperref}
\usepackage{tablists} 
\usepackage{makecell} 
\usepackage{tikz}
\usepackage{hyperref}
\usepackage{pstricks}
\usepackage[titletoc]{appendix}
\usepackage{pstricks-add}
\usepackage{float}
\usepackage{mathrsfs} 
\usepackage{amsthm}
\hypersetup{
	colorlinks=ture,
	linkcolor=blue,
	filecolor=gray,
	urlcolor=blue,
	citecolor=blue}
\numberwithin{equation}{section}
\newtheorem{definition}{Definition}[section]

\allowdisplaybreaks[4]

\begin{document}
	\begin{frontmatter}
		\title{Higher (gauged) Wess–Zumino–Witten terms based on Lie crossed modules} 
	
		\author{Danhua Song}
		\ead{songdh@pku.edu.cn}
		\address{Beijing International Center for Mathematical Research, Peking University, Beijing, China}
		\date{}
		
		\begin{abstract} 
			We derive higher Wess–Zumino–Witten (WZW) and gauged WZW (gWZW) terms within strict higher Chern–Simons (CS) gauge theory. Starting from the Cartan homotopy formula, we obtain the $(2n+2)$-dimensional higher CS forms and transgression forms for strict Lie 2-groups presented by Lie crossed modules. Given two 2-connections related by a higher gauge transformation, higher transgression forms yield canonical higher WZW and gWZW terms.  We prove that, for the symmetric invariant polynomial associated with differential crossed modules, the pure-gauge higher WZW term vanishes identically, whereas the higher gWZW term is exact.  Consequently, the higher CS action is higher-gauge invariant on closed manifolds, and on manifolds with boundary all gauge dependence is encoded in boundary terms.
		\end{abstract}
		
	\end{frontmatter}
	%
	

		\section{Introduction}
		Wess–Zumino–Witten (WZW) terms and their gauged extensions (gWZW terms) \cite{W-Z,Witten,W1} were originally introduced as effective interactions in nuclear and particle physics. In ordinary gauge theory, WZW terms arise  from the bulk--boundary descent associated with Chern–Weil forms and Chern–Simons (CS) functionals, and can be organized conveniently using transgression actions \cite{Bertlmann}. From this perspective, gWZW terms can be viewed as transgression forms evaluated on gauge-equivalent connections, a viewpoint that has been fruitfully exploited in gravitational contexts \cite{ASJ1,PPS,SFNG,PPO}. 
		Whereas a transgression action is strictly gauge invariant, a CS functional is only quasi-invariant: under a gauge transformation its variation reduces to a boundary term, whose part depending only on the gauge transformation accounts for the familiar global features of WZW systems \cite{FEP}. 
	
A natural question is whether an analogous bulk--boundary mechanism persists in higher gauge theory \cite{Baez.2010,JBUS,FH,Baez2005HigherGT}, where gauge symmetry is encoded by higher groups and higher CS theory exist \cite{DHS-3, 	Zucchini-2014-1,Zucchini-2016,R.Z4,R.Z5}. Specifically, does higher transgression produce higher analogues of WZW and gWZW functionals, with a decomposition paralleling the standard CS-transgression relation? In four dimensions, related boundary structures have been obtained from different viewpoints: Refs.~\cite{HC-2024,Vicedo} extract them by an explicit computation of the gauge variation of the higher CS functional, whereas Ref.~\cite{R.Z5} derives closely related terms in a derived framework. A common outcome is that,  the gauge variation is an total boundary term, i.e.\ violation to gauge noninvariance of the higher CS theory is completely holographic. However, a systematic transgression-based derivation, providing uniform control of gauge (non)invariance and boundary reduction, together with an extension of this picture beyond four dimensions to higher CS theories in arbitrary even dimension \cite{DHS-4,DHS-5}, remains underdeveloped.
	
	In this work we address this problem in strict higher gauge theory.  We consider strict Lie $2$-groups presented by Lie crossed modules and work infinitesimally with differential crossed modules and $2$-connections.  Using the Cartan homotopy formula applied to pairs of $2$-connections, we derive explicit higher transgression forms and recover higher CS forms as the special case in which one $2$-connection is trivial.  Specializing higher transgression to two $2$-connections related by a higher gauge transformation yields two canonical boundary functionals: a would-be higher WZW term depending only on higher gauge data, and the corresponding higher gWZW term, together with an explicit bulk--boundary decomposition.
	
	Our main result is a qualitative rigidity statement: for the symmetric invariant polynomial associated with the differential crossed module used here, the higher WZW term vanishes identically.  Consequently, on closed manifolds the $(2n{+}2)$-dimensional higher CS action considered in this paper is  invariant under higher gauge transformations;  in the presence of a boundary, all gauge dependence is boundary-induced.  Moreover, the induced higher gWZW term reduces to an boundary contribution.  Thus, within the strict crossed-module setting, the transgression-based descent mechanism exists but does not produce a non-trivial pure-gauge global WZW functional; any genuinely global boundary effects must therefore come from relaxing the strict framework or altering the underlying invariant data.
	
	The paper is organized as follows.  Section~\ref{sub1} reviews crossed modules and strict higher gauge theory, fixing conventions for higher connections, curvatures, and gauge transformations.  Section~\ref{sub2} derives higher transgression and higher CS forms from the Cartan homotopy formula.  Section~\ref{sub3} analyzes the higher gauge transformation of the higher CS form and derives the associated higher WZW term, proving its vanishing in our setting.  Section~\ref{sub4} constructs the corresponding higher gWZW term via transgression and shows that it is exact.  Section~\ref{sub5} contains conclusions and an outlook.

	\section{Higher gauge theory}\label{sub1}
	This section collects the minimum background on strict higher gauge theory needed later to formulate higher WZW and gWZW terms. The kinematical data of a higher gauge theory are naturally encoded by a homotopy $L_\infty$-algebra \cite{KS,SYH,J2004}. In this work we restrict to the strict case, i.e.\ $L_2$-algebras (strict Lie $2$-algebras) and the corresponding strict Lie $2$-groups, which are conveniently described by Lie crossed modules and their infinitesimal counterparts (differential crossed modules). For further background, see Refs.~\cite{Baez.2010,FH,Baez2005HigherGT,JBUS}.
	
\begin{definition}[\textbf{Lie crossed module}]
	A crossed module $\mathbb{G}=(H,G;\bar{\alpha},\bar{\vartriangleright})$ consists of a group morphism $\bar{\alpha}\colon H\to G$ together with a left action $\bar{\vartriangleright}$ of $G$ on $H$ by automorphisms, such that
	\begin{align*}
		\bar{\alpha}(g\bar{\vartriangleright} h)= g\bar{\alpha}(h)g^{-1}, \qquad
		\bar{\alpha}(h)\bar{\vartriangleright}  h'=h h' h^{-1},
	\end{align*}
	for all $g\in G$ and $h,h'\in H$. If $G$ and $H$ are Lie groups, $\bar{\alpha}$ is smooth, and the action $\bar{\vartriangleright}$ is smooth, then $\mathbb{G}$ is called a \emph{Lie crossed module}.
\end{definition}

The infinitesimal approximation of a Lie $2$-group yields a Lie $2$-algebra. In the strict case this is a $2$-term $L_\infty$-algebra, equivalently a differential crossed module.
\begin{definition}[\textbf{Differential crossed module}]
	A differential crossed module $\mathbb{X}=(\mathcal{h},\mathcal{g};\alpha,\vartriangleright)$ consists of a Lie algebra morphism $\alpha\colon\mathcal{h}\to\mathcal{g}$ together with a left action $\vartriangleright$ of $\mathcal{g}$ on the underlying vector space of $\mathcal{h}$, such that
	\begin{itemize}
		\item[1.] For any $X\in\mathcal{g}$, the map $Y\mapsto X\vartriangleright Y$ is a derivation of $\mathcal{h}$:
		\begin{equation}\label{eq3}
			X\vartriangleright [Y_1,Y_2]=[X\vartriangleright Y_1,Y_2]+[Y_1,X\vartriangleright Y_2],
			\qquad
			\forall X\in\mathcal{g},\ \forall Y_1,Y_2\in\mathcal{h}.
		\end{equation}
		
		\item[2.] The induced map $\mathcal{g}\to \mathrm{Der}(\mathcal{h})$ is a Lie algebra morphism:
		\begin{equation}\label{eq2}
			[X_1,X_2]\vartriangleright Y=
			X_1\vartriangleright(X_2\vartriangleright Y)-X_2\vartriangleright(X_1\vartriangleright Y),
			\qquad
			\forall X_1,X_2\in\mathcal{g},\ \forall Y\in\mathcal{h}.
		\end{equation}
		
		\item[3.] Equivariance of $\alpha$:
		\begin{equation}\label{equivariance}
			\alpha(X\vartriangleright Y)=[X,\alpha(Y)],
			\qquad
			\forall X\in\mathcal{g},\ \forall Y\in\mathcal{h}.
		\end{equation}
		
		\item[4.] Peiffer identity:
		\begin{equation}\label{Peiffer}
			\alpha(Y)\vartriangleright Y'=[Y,Y'],
			\qquad
			\forall Y,Y'\in\mathcal{h}.
		\end{equation}
	\end{itemize}
\end{definition}

Since higher gauge fields are differential forms valued in the relevant Lie algebras, we fix notation for algebra-valued forms and their components.
Let $\Lambda^k(M,\mathcal{g})$ denote the $C^\infty(M)$-module of $\mathcal{g}$-valued differential $k$-forms on a manifold $M$.
For $A\in\Lambda^k(M,\mathcal{g})$, write $A=\sum_a A^a X_a$, where $A^a$ are scalar $k$-forms and $\{X_a\}$ is a basis of $\mathcal{g}$.
Throughout, we assume that $\mathcal{g}$ is a matrix Lie algebra, so that $[X,X']=XX'-X'X$ for $X,X'\in\mathcal{g}$.
We define
\begin{align*}
	dA &:= \sum_a (dA^a)\,X_a,
	\qquad
	A_1\wedge A_2 := \sum_{a,b} A_1^a\wedge A_2^b\,X_aX_b,\\
	[A_1,A_2] &:= \sum_{a,b} A_1^a\wedge A_2^b\,[X_a,X_b],
\end{align*}
for $A_1=\sum_a A_1^a X_a\in\Lambda^{k_1}(M,\mathcal{g})$ and $A_2=\sum_a A_2^a X_a\in\Lambda^{k_2}(M,\mathcal{g})$.
Then
\begin{align*}
	[A_1,A_2]=A_1\wedge A_2-(-1)^{k_1k_2}A_2\wedge A_1.
\end{align*}
The same conventions apply to $\mathcal{h}$-valued forms.
Moreover, for $B=\sum_b B^b Y_b\in\Lambda^{t}(M,\mathcal{h})$ with $Y_b\in\mathcal{h}$, we set
\begin{align*}
	\alpha(B):=\sum_b B^b\,\alpha(Y_b),
	\qquad
	A\vartriangleright B:=\sum_{a,b} A^a\wedge B^b\,(X_a\vartriangleright Y_b).
\end{align*}

Given a $\mathcal{g}$-valued connection $1$-form $A$, we define the covariant derivative $D$ acting on a $\mathcal{g}$-valued form $A'$ and an $\mathcal{h}$-valued form $B$ by
\begin{align*}
	DA' := dA' + [A,A'],
	\qquad
	DB : = dB  + A\vartriangleright B.
\end{align*}

We now recall the basic fields of higher gauge theory, namely $2$-connections.
On a manifold $M$, a (local) $2$-connection is given by a $\mathcal{g}$-valued $1$-form $A\in\Lambda^1(M,\mathcal{g})$ and an $\mathcal{h}$-valued $2$-form $B\in\Lambda^2(M,\mathcal{h})$.
The corresponding fake curvature and $2$-curvature are
\begin{align*}
	\mathcal{F}=dA+\dfrac{1}{2}[A,A]-\alpha(B),
	\qquad
	\mathcal{G}=dB+A\vartriangleright B,
\end{align*}
and they satisfy the $2$-Bianchi identities
\begin{align*}
	d\mathcal{F}+[A,\mathcal{F}]+\alpha(\mathcal{G})=0,\\
	d\mathcal{G}+A\vartriangleright\mathcal{G}-\mathcal{F}\vartriangleright B=0.
\end{align*}
A $2$-connection $(A,B)$ is called \emph{fake-flat} if $\mathcal{F}=0$, and \emph{$2$-flat} if $\mathcal{G}=0$.

A $2$-gauge transformation of a $2$-connection $(A,B)$ is given by
\begin{align}
A^{g,\phi}&=g^{-1}Ag+g^{-1}dg+\alpha(\phi),\label{2gt1}\\
	B^{g,\phi}&=g^{-1}\vartriangleright B+d\phi+	A^{g,\phi}\vartriangleright\phi-\phi\wedge\phi,\label{2gt2}
\end{align}
where $g\in G$ and $\phi\in\Lambda^1(M,\mathcal{h})$.
This is the local transformation law for $2$-connections on $2$-bundles when passing between coordinate neighborhoods.
Gauge transformations form a semidirect product group with multiplication
\begin{equation*}
	(g,\phi)(g',\phi')=(gg',\,g\vartriangleright\phi'+\phi).
\end{equation*}
Under \eqref{2gt1}–\eqref{2gt2}, the fake curvature and $2$-curvature transform as
\begin{align}
		\mathcal{F}^{g,\phi}&=g^{-1}\mathcal{F}g, \label{cur1}
	\\
	\mathcal{G}^{g,\phi}&=g^{-1}\vartriangleright\mathcal{G}+	\mathcal{F}^{g,\phi} \vartriangleright\phi.  \label{cur2}
\end{align}

\section{Higher transgression forms and Cartan homotopy formula}\label{sub2}
In \cite{DHS-5}, higher transgression forms and higher CS forms, together with their relation, were derived from the extended Cartan homotopy formula. Since the ordinary Cartan homotopy formula is a special case thereof, it already suffices to obtain these results. We therefore work directly with the Cartan homotopy formula and re-derive the higher transgression forms (and the associated higher CS forms) in a form tailored for the subsequent construction of higher WZW and gWZW terms. 

Let $(A_t,B_t)$ be the linear interpolation between two $2$-connections $(A_0,B_0)$ and $(A_1,B_1)$,
\begin{align*}
	A_t&=A_0+t\Theta, \qquad \Theta:=A_1-A_0,\\
	B_t&=B_0+t\Phi, \qquad \Phi:=B_1-B_0,
\end{align*}
with $t\in[0,1]$. The corresponding curvatures are
\begin{align*}
	\mathcal{F}_t&
	=\mathcal{F}_0+t\big(d\Theta+[A_0,\Theta]-\alpha(\Phi)\big)+t^2\Theta^2,\\
	\mathcal{G}_t&
	=\mathcal{G}_0+t\big(d\Phi+\Theta\vartriangleright B_0+A_0\vartriangleright\Phi\big)+t^2\Theta\vartriangleright\Phi.
\end{align*}
Differentiating with respect to $t$ yields
\begin{align*}
	\dfrac{\partial}{\partial t}\mathcal{F}_t
	= D_t\Theta-\alpha(\Phi),
	\qquad
	\dfrac{\partial}{\partial t}\mathcal{G}_t
	= D_t\Phi+\Theta\vartriangleright B_t,
\end{align*}
where $D_t\Theta:=d\Theta+[A_t,\Theta]$ and $D_t\Phi:=d\Phi+A_t\vartriangleright\Phi$.

Introduce the homotopy derivation operator $l_t$ by
\begin{align*}
	l_tA_t&=l_tB_t=0,\\
	l_t\mathcal{F}_t&=d_tA_t=dt\,\dfrac{\partial}{\partial t}A_t=dt\,\Theta,\\
	l_t\mathcal{G}_t&=d_tB_t=dt\,\dfrac{\partial}{\partial t}B_t=dt\,\Phi,
\end{align*}
and extend it as a graded derivation on polynomials in $A_t$, $B_t$, $\mathcal{F}_t$, and $\mathcal{G}_t$.
A direct computation gives
\begin{align*}
	(l_td+dl_t)A_t
	&=l_tdA_t
	=l_t\big(\mathcal{F}_t-A_t^2+\alpha(B_t)\big)
	=dt\,\Theta,\\
	(l_td+dl_t)B_t
	&=l_tdB_t
	=l_t\big(\mathcal{G}_t-A_t\vartriangleright B_t\big)
	=dt\,\Phi,\\
	(l_td+dl_t)\mathcal{F}_t
	&=l_td\mathcal{F}_t+dl_t\mathcal{F}_t
	=l_t\big([\mathcal{F}_t,A_t]-\alpha(\mathcal{G}_t)\big)+dl_t\mathcal{F}_t\\
	&=dt\big(D_t\Theta-\alpha(\Phi)\big)
	=dt\,\dfrac{\partial}{\partial t}\mathcal{F}_t,\\
	(l_td+dl_t)\mathcal{G}_t
	&=l_td\mathcal{G}_t+dl_t\mathcal{G}_t
	=l_t\big(\mathcal{F}_t\vartriangleright B_t-A_t\vartriangleright\mathcal{G}_t\big)+dl_t\mathcal{G}_t\\
	&=dt\big(D_t\Phi+\Theta\vartriangleright B_t\big)
	=dt\,\dfrac{\partial}{\partial t}\mathcal{G}_t.
\end{align*}
Consequently, for any polynomial $S$ in $A_t$, $B_t$, $\mathcal{F}_t$, and $\mathcal{G}_t$, one has
\begin{align}\label{poly}
	(l_td+dl_t)S(A_t,B_t,\mathcal{F}_t,\mathcal{G}_t)
	=d_tS(A_t,B_t,\mathcal{F}_t,\mathcal{G}_t)
	=dt\,\dfrac{\partial}{\partial t}S(A_t,B_t,\mathcal{F}_t,\mathcal{G}_t).
\end{align}

Define the homotopy operator $k_{01}$ as the $t$-integrated version of $l_t$,
\begin{align*}
	k_{01}:=\int_0^1 l_t .
\end{align*}
Integrating \eqref{poly} from $t=0$ to $t=1$ yields the \textbf{Cartan homotopy formula} \cite{DHS-5}
\begin{equation}\label{CHF}
	S(A_1,B_1,\mathcal{F}_1,\mathcal{G}_1)-S(A_0,B_0,\mathcal{F}_0,\mathcal{G}_0)
	=(k_{01}d+dk_{01})S(A_t,B_t,\mathcal{F}_t,\mathcal{G}_t).
\end{equation}

We now specialize to the invariant polynomial
\begin{equation*}
	P_{2n+3}(\mathcal{F}_t,\mathcal{G}_t):=\langle \mathcal{F}_t^{n},\mathcal{G}_t\rangle_{\mathcal{g}\mathcal{h}},
\end{equation*}
where $\langle\cdots,\cdot\rangle_{\mathcal{g}\mathcal{h}}$ denotes an invariant pairing between $\mathcal{g}^{\otimes n}$ and $\mathcal{h}$ (see \ref{app1}).
This form is closed,
\begin{equation*}
	d\langle \mathcal{F}_t^{n},\mathcal{G}_t\rangle_{\mathcal{g}\mathcal{h}}=0.
\end{equation*}
Substituting $S=P_{2n+3}$ into \eqref{CHF} gives
\begin{equation*}
	P_{2n+3}(\mathcal{F}_1,\mathcal{G}_1)-P_{2n+3}(\mathcal{F}_0,\mathcal{G}_0)
	=dk_{01}P_{2n+3}(\mathcal{F}_t,\mathcal{G}_t),
\end{equation*}
or, equivalently, the \textbf{higher Chern–Weil theorem}
\begin{equation}\label{CWT}
	\langle \mathcal{F}_1^{n},\mathcal{G}_1\rangle_{\mathcal{g}\mathcal{h}}
	-\langle \mathcal{F}_0^{n},\mathcal{G}_0\rangle_{\mathcal{g}\mathcal{h}}
	=dQ_{2n+2}(A_1,B_1;A_0,B_0),
\end{equation}
where the $(2n+2)$-form $Q_{2n+2}(A_1,B_1;A_0,B_0)$ is defined by
\begin{align*}
	Q_{2n+2}(A_1,B_1;A_0,B_0):=k_{01}P_{2n+3}(\mathcal{F}_t,\mathcal{G}_t).
\end{align*}
Using $l_t\mathcal{F}_t=dt\,\Theta$ and $l_t\mathcal{G}_t=dt\,\Phi$, we obtain the \textbf{higher transgression form} given by
\begin{align}\label{2transg}
	Q_{2n+2}(A_1,B_1;A_0,B_0)
	&=\int_0^1 l_tP_{2n+3}(\mathcal{F}_t,\mathcal{G}_t)\nonumber\\
	&=n\int_0^1 dt\,\langle \Theta\wedge \mathcal{F}_t^{n-1},\mathcal{G}_t\rangle_{\mathcal{g}\mathcal{h}}
	+\int_0^1 dt\,\langle \mathcal{F}_t^{n},\Phi\rangle_{\mathcal{g}\mathcal{h}}.
\end{align}

 \medskip
 \noindent\textbf{Remark.}
 As in the ordinary case, higher transgression forms yield gauge-invariant actions and generalize higher CS theories (see \cite{DHS-4}). In particular, $(A_1,B_1)$ and $(A_0,B_0)$ may be defined on distinct manifolds $M_1$ and $M_0$ sharing a common boundary $\partial M_1=\partial M_0$. One may then treat both pairs as independent dynamical variables, or regard $(A_0,B_0)$ as a fixed (non-dynamical) background.
 
 \medskip
 Finally, consider the special case $A_0=B_0=0$, and write $A_1=A$, $B_1=B$. In this case the homotopy operator is customarily denoted by
 \begin{equation*}
 	k:=\int_0^1 l_t,
 \end{equation*}
 and \eqref{CWT} reduces to
 \begin{equation*}
 	\langle \mathcal{F}^{n},\mathcal{G}\rangle_{\mathcal{g}\mathcal{h}}
 	=dk\langle \mathcal{F}_t^{n},\mathcal{G}_t\rangle_{\mathcal{g}\mathcal{h}}
 	=dQ_{2n+2}(A,B,\mathcal{F},\mathcal{G}),
 \end{equation*}
 where  $Q_{2n+2}(A,B,\mathcal{F},\mathcal{G})=Q_{2n+2}(A,B;0,0)$ as  given explicitly in~\eqref{2transg}. This defines the \textbf{higher CS form}
 \begin{align}\label{2CS}
 	Q_{2n+2}(A,B,\mathcal{F},\mathcal{G})
 	&=k\langle \mathcal{F}_t^{n},\mathcal{G}_t\rangle_{\mathcal{g}\mathcal{h}}\nonumber\\
 	&=n\int_0^1 dt\,\langle A\wedge \mathcal{F}_t^{n-1},\mathcal{G}_t\rangle_{\mathcal{g}\mathcal{h}}
 	+\int_0^1 dt\,\langle \mathcal{F}_t^{n},B\rangle_{\mathcal{g}\mathcal{h}}.
 \end{align}
 Here we take the homotopy fields to be $A_t=tA$ and $B_t=tB$, so that
 \begin{align*}
 	\mathcal{F}_t
 	&
 	=t\mathcal{F}+(t^2-t)A^2,\\
 	\mathcal{G}_t
 	&
 	=t\mathcal{G}+(t^2-t)A\vartriangleright B.
 \end{align*}

\section{Higher  Wess–Zumino–Witten terms}\label{sub3}
In this section, we study the transformation of the higher CS form under higher gauge transformations and, in analogy with the standard CS and WZW correspondence \cite{PPO}, derive the boundary functional that plays the role of the WZW term in the higher gauge theory. 

We will use the higher gauge transformation laws \eqref{2gt1} and \eqref{2gt2} and denote the transformed fields by $(A^{g,\phi},B^{g,\phi})$.
	For later convenience, we introduce 
	\begin{align*}
		A^{g} &:= g^{-1}Ag+g^{-1}dg,\\
		F(\phi) &:= d\phi+\phi\wedge\phi,
	\end{align*}
and define the auxiliary fields
\begin{align*}
	V&:=dgg^{-1}+g\alpha(\phi)g^{-1},\\
	W&:=g\vartriangleright F(\phi)+(dgg^{-1})\vartriangleright(g\vartriangleright \phi).
\end{align*}
With these abbreviations, \eqref{2gt1} and \eqref{2gt2} can be equivalently written as
	\begin{align*}
		A^{g,\phi} &= A^{g}+\alpha(\phi)=g^{-1}(A+V)g,\\
		B^{g,\phi} &= g^{-1}\vartriangleright B + F(\phi) + A^{g}\vartriangleright\phi=g^{-1}\vartriangleright\big(B+W+A\vartriangleright(g\vartriangleright \phi)\big).
\end{align*}
The corresponding curvatures transform as \eqref{cur1} and \eqref{cur2}.


To analyze the transformation of the higher CS form in section \ref{sub2}, we consider the one-parameter family
\begin{subequations}\label{homo}
		\begin{align}
		A_t^{g,\phi}&=A_t^g+\alpha(\phi)=g^{-1}(A_t+V)g,\\
		B_t^{g,\phi}&=g^{-1}\vartriangleright tB+F(\phi)+A_t^g\vartriangleright \phi
		=g^{-1}\vartriangleright\big(B_t+W+A_t\vartriangleright(g\vartriangleright \phi)\big),
	\end{align}

\end{subequations}
with $A_t=tA$, $B_t=tB$, $t\in[0,1]$, and $A_t^g=g^{-1}A_tg+g^{-1}dg$.
The corresponding homotopic fake curvature is given by
\begin{align*}
	\mathcal{F}_t^{g,\phi}=dA_t^g+\dfrac{1}{2}[A_t^g,A_t^g]-g^{-1}\alpha(B_t)g,
\end{align*}
where we have used \eqref{equivariance}.
Recalling that
\begin{equation*}
	F_t^g=dA_t^g+\dfrac{1}{2}[A_t^g,A_t^g]=g^{-1}F_tg,
	\qquad
	F_t=dA_t+\dfrac{1}{2}[A_t,A_t],
\end{equation*}
we obtain
\begin{equation*}
	\mathcal{F}_t^{g,\phi}=g^{-1}\big(F_t-\alpha(B_t)\big)g=g^{-1}\mathcal{F}_tg.
\end{equation*}

A similar computation for the $2$-curvature yields
\begin{align*}
	\mathcal{G}_t^{g,\phi}
	&=dg^{-1}\vartriangleright B_t+g^{-1}\vartriangleright dB_t
	+[d\phi, \phi]
	+dA_t^g\vartriangleright \phi-A_t^g\vartriangleright d\phi
	+A_t^g\vartriangleright(g^{-1}\vartriangleright B_t)
	+A_t^g\vartriangleright F(\phi)\\
	&\quad
	+A_t^g\vartriangleright(A_t^g\vartriangleright \phi)
	+\alpha(\phi)\vartriangleright(g^{-1}\vartriangleright B_t)
	+\alpha(\phi)\vartriangleright F(\phi)
	+\alpha(\phi)\vartriangleright(A_t^g\vartriangleright \phi).
\end{align*}
Using \eqref{eq2}, we have
\begin{align*}
	A_t^g\vartriangleright(A_t^g\vartriangleright \phi)
	=\dfrac{1}{2}[A_t^g,A_t^g]\vartriangleright \phi.
\end{align*}
Moreover, using \eqref{Peiffer}, it follows that
\begin{align*}
	\alpha(\phi)\vartriangleright(g^{-1}\vartriangleright B_t)
	&=-\alpha(g^{-1}\vartriangleright B_t)\vartriangleright \phi
	=-\big(g^{-1}\alpha(B_t)g\big)\vartriangleright \phi,\\
	\alpha(\phi)\vartriangleright F(\phi)
	&=[\phi,d\phi+\phi\phi]=[\phi,d\phi],\\
	\alpha(\phi)\vartriangleright(A_t^g\vartriangleright \phi)
	&=[\phi,A_t^g\vartriangleright \phi].
\end{align*}
Finally, the identity \eqref{eq3} implies
\begin{align*}
	A_t^g\vartriangleright[\phi,\phi]
	=[A_t^g\vartriangleright \phi,\phi]-[\phi,A_t^g\vartriangleright \phi]
	=-2[\phi,A_t^g\vartriangleright \phi].
\end{align*}
Collecting these relations, we obtain the compact expression
\begin{equation*}
	\mathcal{G}_t^{g,\phi}
	=g^{-1}\vartriangleright \mathcal{G}_t+\mathcal{F}_t^{g,\phi}\vartriangleright \phi,
\end{equation*}
where $\mathcal{F}_t=t\mathcal{F}+(t^2-t)A^2$ and $\mathcal{G}_t=t\mathcal{G}+(t^2-t)A\vartriangleright B$.

The homotopy \eqref{homo} interpolates between the gauge-transformed fields at $t=1$,
\begin{align*}
	A_1^{g,\phi}&=g^{-1}Ag+g^{-1}dg+\alpha(\phi)=g^{-1}(A+V)g,\\
	B_1^{g,\phi}&=g^{-1}\vartriangleright B+F(\phi)+A^g\vartriangleright \phi
	=g^{-1}\vartriangleright\big(B+W+A\vartriangleright(g\vartriangleright \phi)\big),
\end{align*}
and the reference configuration at $t=0$,
\begin{align*}
	A_0^{g,\phi}&=g^{-1}dg+\alpha(\phi)=g^{-1}Vg,\\
	B_0^{g,\phi}&=F(\phi)+(g^{-1}dg)\vartriangleright \phi=g^{-1}\vartriangleright W.
\end{align*}
At the endpoints the curvatures satisfy
\begin{equation*}
	\mathcal{F}_1^{g,\phi}=\mathcal{F}^{g,\phi},\qquad
	\mathcal{G}_1^{g,\phi}=\mathcal{G}^{g,\phi},
\end{equation*}
while at $t=0$ one has
\begin{equation*}
	\mathcal{F}_0^{g,\phi}=0,\qquad
	\mathcal{G}_0^{g,\phi}=0.
\end{equation*}

Applying the Cartan homotopy formula \eqref{CHF} to the higher CS form \eqref{2CS} along the family $(A_t^{g,\phi},B_t^{g,\phi},$ $\mathcal{F}_t^{g,\phi},\mathcal{G}_t^{g,\phi})$, we obtain
\begin{align}\label{CHF-CS}
	&Q_{2n+2}(A_1^{g,\phi},B_1^{g,\phi},\mathcal{F}_1^{g,\phi},\mathcal{G}_1^{g,\phi})
	-Q_{2n+2}(A_0^{g,\phi},B_0^{g,\phi},\mathcal{F}_0^{g,\phi},\mathcal{G}_0^{g,\phi})\nonumber\\
	&\qquad=
	Q_{2n+2}(A^{g,\phi},B^{g,\phi},\mathcal{F}^{g,\phi},\mathcal{G}^{g,\phi})
	-Q_{2n+2}(g^{-1}Vg,g^{-1}\vartriangleright W,0,0)\nonumber\\
	&\qquad=(k_{01}d+dk_{01})\,Q_{2n+2}(A_t^{g,\phi},B_t^{g,\phi},\mathcal{F}_t^{g,\phi},\mathcal{G}_t^{g,\phi}).
\end{align}

From the definition \eqref{2CS}, the higher CS form after a higher gauge transformation admits the explicit expression
\begin{align*}
	Q_{2n+2}(A^{g, \phi}, B^{g, \phi}, \mathcal{F}^{g, \phi}, \mathcal{G}^{g, \phi})=\int_{0}^{1}ds\Big\{n\langle A^{g, \phi} \wedge (\mathcal{F}_s^{g, \phi})^{n-1}, \mathcal{G}_s^{g, \phi}\rangle_{\mathcal{g}\mathcal{h}}+  \langle (\mathcal{F}_s^{g, \phi})^n, B^{g, \phi}\rangle_{\mathcal{g}\mathcal{h}}\Big\}
\end{align*}
where 
\begin{align*}
\mathcal{F}_s^{g, \phi}&=s\mathcal{F}^{g, \phi}+(s^2-s)(A^{g, \phi})^2=g^{-1}\hat{\mathcal{F}_s}g,\\
\mathcal{G}_s^{g, \phi}&=s\mathcal{G}^{g, \phi}+(s^2-s)A^{g, \phi}\vartriangleright B^{g, \phi}=g^{-1}\vartriangleright \hat{\mathcal{G}_s}
\end{align*}
with 
\begin{align*}
\hat{\mathcal{F}_s}&=s\mathcal{F}+(s^2-s)(A+V)^2,\\
	\hat{\mathcal{G}_s}&=s\big(\mathcal{G}+\mathcal{F}\vartriangleright(g\vartriangleright \phi)\big)+(s^2-s)(A+V)\vartriangleright\big(B+W+A\vartriangleright(g\vartriangleright \phi)\big).
\end{align*}
Using the invariance of the pairing, we obtain
\begin{align*}
	&Q_{2n+2}(A^{g, \phi}, B^{g, \phi}, \mathcal{F}^{g, \phi}, \mathcal{G}^{g, \phi})\nonumber\\
	=&\int_{0}^{1}ds\Big\{ n \langle g^{-1}(A+V)g(g^{-1} \hat{\mathcal{F}_s}g)^{n-1}, g^{-1}\vartriangleright \hat{\mathcal{G}_s}\rangle_{\mathcal{g}\mathcal{h}}+\langle (g^{-1} \hat{\mathcal{F}_s}g)^{n}, g^{-1}\vartriangleright \big(B+W+A\vartriangleright(g\vartriangleright \phi)\big)\rangle_{\mathcal{g}\mathcal{h}}\Big\}\nonumber\\
	=&\int_{0}^{1}ds\Big\{ n \langle (A+V) \hat{\mathcal{F}_s}^{n-1},  \hat{\mathcal{G}_s}\rangle_{\mathcal{g}\mathcal{h}} +\langle  \hat{\mathcal{F}_s}^{n},  B+W+A\vartriangleright(g\vartriangleright \phi)\rangle_{\mathcal{g}\mathcal{h}}\Big\}.
\end{align*}

Analogously, for the homotopic family one has
\begin{align*}
	Q_{2n+2}(A_t^{g, \phi}, B_t^{g, \phi}, \mathcal{F}_t^{g, \phi}, \mathcal{G}_t^{g, \phi})=\int_{0}^{1}ds\Big\{n\langle A_t^{g, \phi} \wedge (\mathcal{F}_{ts}^{g, \phi})^{n-1}, \mathcal{G}_{ts}^{g, \phi}\rangle_{\mathcal{g}\mathcal{h}}+  \langle (\mathcal{F}_{ts}^{g, \phi})^n, B_t^{g, \phi}\rangle_{\mathcal{g}\mathcal{h}}\Big\},
\end{align*}
where
\begin{align*}
	\mathcal{F}_{ts}^{g, \phi}&=s\mathcal{F}_t^{g, \phi}+(s^2-s)(A_t^{g, \phi})^2=g^{-1}\hat{\mathcal{F}_{ts}}g,\\
	\mathcal{G}_{ts}^{g, \phi}&=s\mathcal{G}_t^{g, \phi}+(s^2-s)A_t^{g, \phi}\vartriangleright B_t^{g, \phi}=g^{-1}\vartriangleright \hat{\mathcal{G}_{ts}},
\end{align*}
with 
\begin{align*}
	\hat{\mathcal{F}_{ts}}&=s\mathcal{F}_t+(s^2-s)(A_t+V)^2,\\
	\hat{\mathcal{G}_{ts}}&=s\big(\mathcal{G}_t+\mathcal{F}_t\vartriangleright(g\vartriangleright \phi)\big)+(s^2-s)(A_t+V)\vartriangleright\big(B_t+W+A_t\vartriangleright(g\vartriangleright \phi)\big).
\end{align*}
Substituting these expressions and again using the invariance of the pairing yields
\begin{align*}
		Q_{2n+2}(A_t^{g, \phi}, B_t^{g, \phi}, \mathcal{F}_t^{g, \phi}, \mathcal{G}_t^{g, \phi})=\int_{0}^{1}ds\Big\{ n \langle (A_t+V) \hat{\mathcal{F}_{ts}}^{n-1},  \hat{\mathcal{G}_{ts}}\rangle_{\mathcal{g}\mathcal{h}} +\langle  \hat{\mathcal{F}_{ts}}^{n},  B_t+W+A_t\vartriangleright(g\vartriangleright \phi)\rangle_{\mathcal{g}\mathcal{h}}\Big\}.
\end{align*}

To evaluate the right-hand side of \eqref{CHF-CS}, we recall the general identity for the exterior derivative of the higher CS form,
\begin{align*}
	dQ_{2n+2}(A_t^{g, \phi}, B_t^{g, \phi}, \mathcal{F}_t^{g, \phi}, \mathcal{G}_t^{g, \phi})=\langle (\mathcal{F}_t^{g, \phi})^n, \mathcal{G}_t^{g, \phi}\rangle_{\mathcal{g}\mathcal{h}}=\langle (\mathcal{F}_t)^n, \mathcal{G}_t\rangle_{\mathcal{g}\mathcal{h}}.
\end{align*}
Consequently,
\begin{align*}
	k_{01}dQ_{2n+2}(A_t^{g, \phi}, B_t^{g, \phi}, \mathcal{F}_t^{g, \phi}, \mathcal{G}_t^{g, \phi})&=k_{01}\langle (\mathcal{F}_t)^n, \mathcal{G}_t\rangle_{\mathcal{g}\mathcal{h}}=\int_{0}^{1}l_t \langle (\mathcal{F}_t)^n, \mathcal{G}_t\rangle_{\mathcal{g}\mathcal{h}}\nonumber\\
	&=Q_{2n+2}(A, B, \mathcal{F}, \mathcal{G}),
\end{align*}
where the last equality follows from \eqref{2CS}.
Besides, define the $(2n+1)$-form
\begin{equation*}
	\alpha_{2n+1}:=k_{01} Q_{2n+2}(A_t^{g, \phi}, B_t^{g, \phi}, \mathcal{F}_t^{g, \phi}, \mathcal{G}_t^{g, \phi})=\int_{0}^{1}l_t Q_{2n+2}(A_t^{g, \phi}, B_t^{g, \phi}, \mathcal{F}_t^{g, \phi}, \mathcal{G}_t^{g, \phi}).
\end{equation*}

Then the Cartan homotopy formula \eqref{CHF-CS} simplifies to
\begin{equation}\label{tr1}
	Q_{2n+2}(A^{g, \phi}, B^{g, \phi}, \mathcal{F}^{g, \phi}, \mathcal{G}^{g, \phi})-Q_{2n+2}(A, B, \mathcal{F}, \mathcal{G})=Q_{2n+2}(g^{-1}Vg, g^{-1}\vartriangleright W, 0, 0) +d\alpha_{2n+1}.
\end{equation}
We identify the first term on the right-hand side of \eqref{tr1} as the \textbf{higher WZW term}.

A direct evaluation from \eqref{2CS} gives
\begin{align*}
	Q_{2n+2}(g^{-1}Vg, g^{-1}\vartriangleright W, 0, 0)=\int_{0}^{1}dt\Big\{n\langle g^{-1}Vg (\bar{\mathcal{F}}_t)^{n-1},\bar{\mathcal{G}}_t \rangle_{\mathcal{g}\mathcal{h}}+\langle (\bar{\mathcal{F}}_t)^{n},g^{-1}\vartriangleright W\rangle_{\mathcal{g}\mathcal{h}}
	\Big\},
\end{align*}
where in the present case
\begin{align*}
\bar{\mathcal{F}}_t=(t^2-t)g^{-1}V^2g, \qquad 
\bar{\mathcal{G}}_t=(t^2-t)(g^{-1}Vg)\vartriangleright (g^{-1}\vartriangleright W).
\end{align*}
Substituting these expressions and using the invariance of the pairing, we obtain
\begin{align}\label{hWZW1}
	Q_{2n+2}(g^{-1}Vg, g^{-1}\vartriangleright W, 0, 0)=&\int_{0}^{1}dt\Big\{n\langle g^{-1}Vg ((t^2-t)g^{-1}V^2g)^{n-1},(t^2-t)(g^{-1}Vg)\vartriangleright (g^{-1}\vartriangleright W) \rangle_{\mathcal{g}\mathcal{h}}\nonumber\\
	&+\langle ((t^2-t)g^{-1}V^2g)^{n},g^{-1}\vartriangleright W\rangle_{\mathcal{g}\mathcal{h}}
	\Big\}\nonumber\\
	=&\Big\{ n\langle V (V^2)^{n-1}, V\vartriangleright W\rangle_{\mathcal{g}\mathcal{h}}+\langle (V^2)^{n}, W\rangle_{\mathcal{g}\mathcal{h}}\Big\}
	\int_{0}^{1}(t^2-t)^ndt.
\end{align}
Employing \eqref{sy2} and \eqref{sy4}, we further find 
\begin{equation*}
	\langle V (V^2)^{n-1}, V\vartriangleright W\rangle_{\mathcal{g}\mathcal{h}}=2\langle (V^2)^{n}, W\rangle_{\mathcal{g}\mathcal{h}}. 
\end{equation*}
The remaining integral can be evaluated using the Beta function:
\begin{equation*}
	B(m+1, n+1)=\int_{0}^{1}dt t^m(1-t)^n=\dfrac{m!n!}{(m+n+1)!},
\end{equation*}
Indeed,
\begin{equation*}
	\int_{0}^{1}(t^2-t)^n\,dt
	=(-1)^n\int_{0}^{1}t^n(1-t)^n\,dt
	=(-1)^n B(n+1,n+1)
	=(-1)^n\frac{n!\,n!}{(2n+1)!}.
\end{equation*}

Collecting these results, the higher WZW term \eqref{hWZW1} simplifies to
\begin{equation*}
	Q_{2n+2}(g^{-1}Vg, g^{-1}\vartriangleright W, 0, 0)=(-1)^n\dfrac{n!n!}{(2n)!}\langle V^{2n}, W\rangle_{\mathcal{g}\mathcal{h}},
\end{equation*}
which is independent of the higher gauge-field components $A$ and $B$, and depends only on the higher gauge-transformation parameters $g$ and $\phi$.
Therefore, $Q_{2n+2}(g^{-1}Vg, g^{-1}\vartriangleright W, 0, 0)$ provides a higher analogue of the familiar WZW term \cite{PPO}.
The derivation above closely parallels the standard construction of the WZW term in CS theory, with the main difference that the computation is performed in the framework of higher gauge theory.

However, since the multilinear symmetric invariant polynomial for the differential crossed module $(\mathcal{h}, \mathcal{g}; \alpha, \vartriangleright)$ is symmetric in the $\mathcal{g}$-entries in \eqref{sy4}, the higher WZW term vanishes:
\begin{equation}\label{hWZW}
	Q_{2n+2}(g^{-1}Vg, g^{-1}\vartriangleright W, 0, 0)=0.
\end{equation}
Consequently, \eqref{tr1} simplifies to
\begin{equation*}
	Q_{2n+2}(A^{g, \phi}, B^{g, \phi}, \mathcal{F}^{g, \phi}, \mathcal{G}^{g, \phi})-Q_{2n+2}(A, B, \mathcal{F}, \mathcal{G})=d\alpha_{2n+1}.
\end{equation*}

Although the $(2n+2)$-dimensional higher CS model can be formulated so as to parallel, in many respects, the familiar $(2n+1)$-dimensional CS theory, its gauge-invariance properties are remarkably different.
In particular, on a manifold without boundary the $(2n+2)$-dimensional higher CS action is \emph{strictly} gauge invariant.
Accordingly, any gauge non-invariance is entirely boundary-induced: when a boundary is present, the gauge variation reduces to a boundary term.
These statements generalize the conclusions of \cite{R.Z5} for the $4$D higher CS theory ($n=1$) to arbitrary even dimensions.

\section{Higher gauged Wess–Zumino–Witten terms}\label{sub4}
Motivated by the standard construction of gWZW terms   \cite{PPS,SFNG,PPO}, we introduce higher gWZW terms formulated in terms of Lie crossed modules. This point is not discussed in \cite{R.Z5} in the context of the four-dimensional  higher CS theory.

Applying the Cartan homotopy formula \eqref{CHF} to the higher CS form \eqref{2CS}, we obtain
\begin{align}\label{eqq}
	Q_{2n+2}(A_1, B_1;A_0, B_0)=&Q_{2n+2}(A_1, B_1, \mathcal{F}_1, \mathcal{G}_1)-Q_{2n+2}(A_0, B_0, \mathcal{F}_0, \mathcal{G}_0)\nonumber\\
	&-dk_{01}Q_{2n+2}(A_t, B_t, \mathcal{F}_t, \mathcal{G}_t),
\end{align}
where $A_t=A_0+t(A_1-A_0)$ and $B_t=B_0+t(B_1-B_0)$. 

Define the $(2n+1)$-form
\begin{equation*}
	\mathbf{B}_{2n+1}=k_{01}Q_{2n+2}(A_t, B_t, \mathcal{F}_t, \mathcal{G}_t).
\end{equation*}
Then \eqref{eqq} can be rewritten as
\begin{equation}\label{eq1}
		Q_{2n+2}(A_1, B_1;A_0, B_0)=Q_{2n+2}(A_1, B_1, \mathcal{F}_1, \mathcal{G}_1)-Q_{2n+2}(A_0, B_0, \mathcal{F}_0, \mathcal{G}_0)-d \mathbf{B}_{2n+1},
\end{equation}
where $\mathbf{B}_{2n+1}$ admits the explicit expression
\begin{equation*}
\mathbf{B}_{2n+1}=-\int_{0}^{1}dt \int_{0}^{1}ds ns \Big\{ (n-1)\langle A_t \wedge \Theta \wedge \mathcal{F}^{n-2}_{st}, \mathcal{G}_{st}\rangle_{\mathcal{g}\mathcal{h}} + \langle A_t \wedge \mathcal{F}^{n-1}_{st}, \Phi \rangle_{\mathcal{g}\mathcal{h}} - \langle \Theta \wedge \mathcal{F}^{n-1}_{st}, B_t \rangle_{\mathcal{g}\mathcal{h}} \Big\}.
\end{equation*}
Here we use the notation
\begin{align*}
	A_t&=A_0+t \Theta,\qquad A_{st}=sA_t=sA_0+st \Theta,\qquad \mathcal{F}_{st}=s\mathcal{F}_t+(s^2-s)A_t^2,\\
	B_t&=B_0+t \Phi,\qquad  B_{st}=sB_t= sB_0+st \Phi,\qquad \mathcal{G}_{st}=s\mathcal{G}_t+(s^2-s)A_t\wedge^{\vartriangleright} B_t,
\end{align*}
with $\Theta= A_1-A_0$ and $\Phi= B_1-B_0$.

We now specialize to the case where $(A_1,B_1)$ is obtained from $(A_0,B_0)$ by a higher gauge transformation. More precisely, we take

	\begin{align*}
		A_1&=A^{g, \phi}=g^{-1}Ag +g^{-1}dg +\alpha(\phi),  \qquad \qquad 	A_0=A,\\
		B_1&=B^{g,\phi}=g^{-1}\vartriangleright B +F(\phi)+A^g\vartriangleright \phi,
	  \qquad 
		B_0=B.
	\end{align*}

Substituting these relations into Eq.~\eqref{eq1}, we obtain
\begin{align}\label{tr2}
	Q_{2n+2}(A^{g, \phi}, B^{g, \phi};A, B)=&Q_{2n+2}(A^{g, \phi}, B^{g, \phi}, \mathcal{F}^{g, \phi}, \mathcal{G}^{g, \phi})-Q_{2n+2}(A, B, \mathcal{F}, \mathcal{G})\nonumber\\
	&-d \mathbf{B}_{2n+1}(A^{g, \phi}, B^{g, \phi}; A, B).
\end{align}

Comparing \eqref{tr1} and \eqref{tr2}, we find that for gauge-equivalent $2$-connections the corresponding higher transgression form is encoded by the higher \textbf{gWZW} term:
\begin{align*}
	Q_{2n+2}(A^{g, \phi}, B^{g, \phi};A, B)=&Q_{2n+2}(g^{-1}Vg, g^{-1}\vartriangleright W, 0, 0) +d\alpha_{2n+1}\nonumber\\
	& -d \mathbf{B}_{2n+1}(A^{g, \phi}, B^{g, \phi}; A, B),
\end{align*}
which may be viewed as a higher analogue of the familiar gWZW term \cite{PPO}. Moreover, since the higher WZW term vanishes, cf.\ \eqref{hWZW}, the higher gWZW term is exact:
\begin{align*}
	Q_{2n+2}(A^{g, \phi}, B^{g, \phi};A, B)=d\big(\alpha_{2n+1}
	 - \mathbf{B}_{2n+1}(A^{g, \phi}, B^{g, \phi}; A, B)\big).
\end{align*}

In summary, the higher \textbf{gWZW} term defined above has the same conceptual origin as the ordinary gWZW functional: it arises canonically from the (higher) transgression form via the Cartan homotopy construction and captures the gauge dependence through an exact differential, in direct analogy with the CS and WZW correspondence \cite{PPS,SFNG,PPO}. The crucial difference is that, in the ordinary gauge theory, the gWZW contribution is in general only quasi-invariant and may contain a genuinely global part, which underlies the usual level-quantization mechanism. In the present higher framework, by contrast, the higher WZW term vanishes, and therefore the resulting higher gWZW term is a pure coboundary.

\section{Conclusions and outlook}\label{sub5}
In this paper we have established three main results. First, we extend the use of the Cartan homotopy formula of \cite{FEP1, PRRJ1, PRRJ} to the setting of strict higher gauge theory and, within this framework, re-derive the higher Chern–Weil theorem together with the associated transgression forms and higher CS forms as in \cite{DHS-4}. Second, we introduce a higher analogue of the WZW term and show that, for gauge data encoded by Lie crossed modules, this higher WZW term vanishes. Third, motivated by the construction schemes of the gWZW term in \cite{PPS,SFNG,PPO}, we generalize the gWZW construction to Lie crossed modules and prove that the resulting higher gWZW terms are exact.

The analysis presented here is restricted to the strict (crossed-module) formulation of higher gauge theory. An important direction for future work is to go beyond strictness and to extend the present framework to semistrict or weak higher groups, such as those governed by Lie $2$-algebras and, more generally, $L_\infty$-algebraic structures \cite{	DFUSJS, PRCS}. A central problem is whether an appropriate extension of the Cartan homotopy formula can be formulated in these broader contexts, and whether the corresponding transgression hierarchy survives. If such an extension exists, it becomes natural to ask whether the construction mechanism developed in this work still produces well-defined higher WZW terms, whether they continue to vanish, and whether the associated higher gWZW terms remain exact in the semistrict or weak setting.

\section*{Acknowledgements}
The author thanks the Beijing International Center for Mathematical Sciences (BICMS) for its hospitality and support.

\appendix
\section{Symmetric invariant polynomials}\label{app1}
Motivated by the usual invariant polynomials $\langle\cdots\rangle_{\mathcal{g}}$ on a Lie algebra $\mathcal{g}$, one may introduce a multilinear, symmetric invariant polynomial associated with a differential crossed module $(\mathcal{h},\mathcal{g};\alpha,\vartriangleright)$ \cite{DHS-4,DHS-5},
\begin{align}\label{mp}
	\langle \cdots,\cdot \rangle_{\mathcal{g}\mathcal{h}}:\ \mathcal{g}^{\otimes n}\times\mathcal{h}\longrightarrow\mathbb{R},
\end{align}
which can be viewed as an invariant pairing between $\mathcal{g}^{\otimes n}$ and $\mathcal{h}$.

This pairing is symmetric in its $\mathcal{g}$-entries and satisfies the invariance conditions
\begin{align}
	&\langle X_1\cdots X_i\cdots X_n,\; X\vartriangleright Y\rangle_{\mathcal{g}\mathcal{h}}
	=-\sum_{i=1}^{n}\langle X_1\cdots [X,X_i]\cdots X_n,\; Y\rangle_{\mathcal{g}\mathcal{h}},\label{sy2}\\
	&\langle X_1\cdots \alpha(Y_i)\cdots X_n,\; Y\rangle_{\mathcal{g}\mathcal{h}}
	=\langle X_1\cdots \alpha(Y)\cdots X_n,\; Y_i\rangle_{\mathcal{g}\mathcal{h}}.\nonumber
\end{align}

Symmetry means that for any $1\le i,j\le n$,
\begin{align}\label{sy4}
	\langle X_1\cdots X_i\cdots X_j\cdots X_n,\; Y\rangle_{\mathcal{g}\mathcal{h}}
	=\langle X_1\cdots X_j\cdots X_i\cdots X_n,\; Y\rangle_{\mathcal{g}\mathcal{h}}.
\end{align}
Furthermore, \eqref{sy2} implies $G$-invariance in the integrated (finite) form: for any $g\in G$,
\begin{align*}
	\langle gX_1g^{-1}\cdots gX_ng^{-1},\; g\vartriangleright Y\rangle_{\mathcal{g}\mathcal{h}}
	=\langle X_1\cdots X_n,\; Y\rangle_{\mathcal{g}\mathcal{h}},
\end{align*}
which follows by taking $g$ infinitesimally close to the identity and using \eqref{sy2}.
For $n=1$, \eqref{mp} reduces to a bilinear pairing
$\langle\cdot,\cdot\rangle_{\mathcal{g}\mathcal{h}}:\mathcal{g}\times\mathcal{h}\to\mathbb{R}$,
as considered in \cite{DHS-3,R.Z5}.

\end{document}